\documentclass{iau} 
\usepackage{graphicx}

\def\ls{\mathrel{\hbox{\rlap{\hbox{\lower4pt\hbox{$\sim$}}}\hbox{$<$}}}}
\def\gs{\mathrel{\hbox{\rlap{\hbox{\lower4pt\hbox{$\sim$}}}\hbox{$>$}}}}

\title[Strong-lensing of Gravitational Waves]{Strong-lensing of Gravitational Waves by Galaxy Clusters}

\author[G.\ P.\ Smith et al.]   
       {Graham P. Smith,$\!^{1\dagger}$
        Christopher Berry,$\!^{1,2}$
        Matteo Bianconi,$\!^1$ \\[\affilskip]
        Will M.\ Farr,$\!^{1,2}$ 
        Mathilde Jauzac,$\!^{3,4,5,6}$ 
        Richard Massey,$\!^3$ \\[\affilskip]
        Johan Richard,$\!^7$ 
        Andrew Robertson,$\!^4$
        Keren Sharon,$\!^8$ \\[\affilskip]
        Alberto Vecchio,$\!^{1,2}$
        John Veitch.$\!^9$ \\[\affilskip]}

\affiliation{ 
  $^1$School of Physics and Astronomy, University of Birmingham,
  Birmingham, B15 2TT, England
  \\[\affilskip] 
  $^2$Birmingham Institute of Gravitational Wave Astronomy, University of Birmingham, Birmingham, B15 2TT, England
  \\[\affilskip] 
  $^3$Centre for Extragalactic Astronomy, Department of Physics, Durham University, Durham DH1 3LE, England
  \\[\affilskip] 
  $^4$Institute for Computational Cosmology, Durham University, South Road, Durham DH1 3LE, England
  \\[\affilskip] 
  $^5$Astrophysics and Cosmology Research Unit, School of Mathematical Sciences, University of KwaZulu-Natal, Durban 4041, South Africa
  \\[\affilskip] 
  $^6$Laboratoire d'Astrophysique \'Ecole Polytechnique F\'ed\'erale de Lausanne (EPFL), Observatoire de Sauverny CH-1290 Versoix, Switzerland  
  \\[\affilskip] 
  $^7$CRAL, Observatoire de Lyon, Universit\'e Lyon 1, 9 Avenue Ch. Andr\'e, 69561 Saint Genis Laval Cedex, France
  \\[\affilskip] 
  $^8$Department of Astronomy, University of Michigan, 500 Church
St., Ann Arbor, MI 48109 USA
  \\[\affilskip] 
  $^9$School of Physics and Astronomy, University of Glasgow, G12 8QQ, Scotland
  \\[\affilskip] 
  $^\dagger$Email: {\tt gps@star.sr.bham.ac.uk}
}

\pubyear{2018}
\volume{338}
\jname{Gravitational Waves Astrophysics: Early Results from GW Searches and Electromagnetic Counterparts}
\editors{A.C. Editor, B.D. Editor \& C.E. Editor, eds.}
\begin{document}

\maketitle

\sloppy

\begin{abstract}
  Discovery of strongly-lensed gravitational wave (GW) sources will
  unveil binary compact objects at higher redshifts and lower
  intrinsic luminosities than is possible without lensing.  Such
  systems will yield unprecedented constraints on the mass
  distribution in galaxy clusters, measurements of the polarization of
  GWs, tests of General Relativity, and constraints on the Hubble
  parameter.  Excited by these prospects, and intrigued by the
  presence of so-called ``heavy black holes'' in the early detections
  by LIGO-Virgo, we commenced a search for strongly-lensed GWs and
  possible electromagnetic counterparts in the latter stages of the
  second LIGO observing run (O2).  Here, we summarise our calculation
  of the detection rate of strongly-lensed GWs, describe our review of
  BBH detections from O1, outline our observing strategy in O2,
  summarize our follow-up observations of GW170814, and discuss the
  future prospects of detection.  
\end{abstract}

\firstsection

\section{Strong-lensing of gravitational waves by galaxy clusters}

The detection of strongly-lensed gravitational wave (GW) sources will
create opportunities to explore binary compact objects at higher
redshifts and/or that are intrinsically fainter than those that are
detectable without assistance from lensing.  This greater ``reach'' of
GW detectors that can be achieved with lensing is due to the
gravitational magnification, $\mu$, that boosts the amplitude of the
strain signal ($A$) at an interferometer by a factor $\sqrt{\mu}$ and
the spectral flux density of any accompanying electromagnetic signal
by a factor $\mu$.  In addition to increasing the sensitivity of
observations, the gravitational magnification has important
consequences for the interpretation of the strain and EM signals.  In
particular, there is a degeneracy between the luminosity distance
($D_L$) to the source and the gravitational magnification, such that
$A\propto D_L^{-1}|\mu|^{0.5}$.  Therefore, if it is assumed that
$\mu=1$ for a source that is actually lensed, then $D_L$ and the
redshift of the source ($z$) will be under-estimated.  Moreover, given
that the rest-frame mass of the source goes as $(1+z)^{-1}$, the mass
of the source would be over-estimated if $\mu=1$ is erroneously
assumed (Wang, Stebbins \& Turner 1996).

The early detections of binary black hole (BBH) GW sources by LIGO and
Virgo are intriguing in the context of gravitational lensing because a
large fraction of the reported BHs have rest frame masses larger than
the most massive stellar mass BHs detected thus far (Abbott et
al. 2016a, 2016b, 2017a, 2017b; Farr et al. 2011).  This encourages
speculation that one or more of the sources could have been
gravitationally lensed.

The realisation that a GW source is strongly-lensed would also open up
the possibility to detect the same GW source on more than one occasion
by virtue of the different arrival times along multiple sight lines
that connect an observer with a strongly-lensed source.  Given the
sub-millisecond accuracy of the measurement of the arrival time of a
GW signal, it would be possible to measure the time delay between
different sight lines through a galaxy cluster to a single GW source
to a similar accuracy.  Such accuracy is $\gs8$ orders of magnitude
better than is achievable with multiply-imaged supernovae and quasars
(Fohlmeister et al., 2007; Rodney et al., 2016).  Discovery of one or
more multiply-imaged GW sources would therefore deliver an
unprecedented constraint on the local mass distribution in cluster
lenses, and potentially offer a new and accurate probe of the Hubble
parameter (e.g. Liao et al. 2017).  Moreover, multiple detections of
the same GW are unlikely to be made when the detectors are in the same
orientations relative to the source.  Therefore, observations of a
multiply-imaged GW source should help to further constrain GW
polarizations and potentially achieve new tests of General Relativity,
since we would effectively observe the same signal with a greater
number of detectors (Chatziioannou, Yunes \& Cornish 2012).

\section{The probability of strong-lensing}

Given that the BBH GW sources are located at $z\sim0.1-0.3$ (assuming
$\mu=1$), a magnification factor of $\mu\sim30-300$ is required to
reinterpret the strain signal as originating from a source at (say)
$z\simeq1$ (Smith et al. 2018).  The physical region from which GWs
emerge is $\sim 100\,{\rm km}$ in size.  It is therefore possible for
a GW source to be very closely aligned with the caustic of a
gravitational lens, and thus achieve such a high magnification value
(Ng et al. 2018) -- i.e.\ we can treat GWs as point sources.  In
contrast, galaxies have typical physical sizes of $\sim1-10\,{\rm
  kpc}$, and therefore cannot be treated as point sources at optical
wavelengths.  Hilbert et al. (2008) predicted that the optical depth
to strong-lensing of point sources is dominated by galaxy clusters,
with the optical depth peaking at $M\simeq10^{14}M_\odot$.  This peak,
and the dominance of clusters, is predicted to be more pronounced for
large lens magnifications, i.e.\ $\mu>10$ (Figure 5 of Hilbert et
al.).  These predictions appear to be borne out by observations, in
that the typical lens magnifications suffered by quasars -- a class of
point source, and thus directly relevant to lensing of GWs -- that are
strongly-lensed by clusters is typically $\mu>10$ (Oguri et al. 2010,
2013; Sharon et al. 2017), whilst the magnification of quasars
strongly-lensed by galaxies is typically $\mu<10$ (Inada et al. 2014).
Therefore, the available predictions and observations both point to
galaxy clusters being more relevant than galaxies (Broadhurst et
al. 2018; Li et 2018; Ng et al. 2018) to the discovery of
strongly-lensed GWs with LIGO-Virgo.

There are currently 130 known and spectroscopically confirmed cluster
strong lenses (Smith et al. 2018 and references therein).  It is
important to stress that spectroscopic confirmation of the redshift of
multiply-imaged galaxies seen through these cluster cores is essential
to achieve robust constraints on the gravitational optics (e.g.\ Smith
et al. 2009).  Most of the cluster lenses are at redshifts of
$z\sim0.2-0.5$; at lower redshifts strong-lensing clusters become
increasingly rare due to the small observable cosmic volume, and
higher redshifts await systematic exploration.

The assumption of $\mu=1$ in the LIGO-Virgo analysis of BBH detections
in the first and second observing runs (O1 and O2) rightly reflects
the fact that the probability of strong-lensing is small.  This is due
to a combination of factors including the large magnifications noted
above, and existing constraints on the local BBH merger rate.
Nevertheless, Smith et al. (2018) showed that the probability is
non-zero.  Based on the known population of cluster lenses and the GW
detections from O1 (Abbott et al., 2016b), they estimated a lower
limit on the rate of detectable strongly-lensed GWs of $R_{\rm
  detect}\simeq10^{-5}$ per detector year.  This corresponds to
rejecting the hypothesis that one of the O1 GW detections was
multiply-imaged at $\ls4\sigma$.

\section{Known strong-lensing clusters and LIGO's O1 BBHs}

Smith et al.\ (2018) identified three strong-lensing clusters with
celestial coordinates consistent with the 90\% credible sky
localisations of GW sources detected in LIGO's O1 (Abbott et
al., 2016b), of which none are consistent with GW150914, two are
consistent with GW151226 (MACS\,J0140.0$-$0555 and
MACS\,J1311.0$-$0310) and one is consistent with LVT151012
(RCS\,0224$-$0002).  Following Jauzac et al. (2016) we used detailed
parameterized models of the mass distribution in each of these cluster
lenses to explore whether and when another appearance of GW151226 and
LVT151012 would be detectable, if they have been strongly-lensed.  In
summary, the next appearance will be detectable at the current
LIGO-Virgo sensitivity in roughly half of the cases, and on a
timescale of up to 3 years from the initial detection.  The
``re-detection rate'' of $\sim50\%$ arises because the relevant
strong-lensing configurations typically involve one caustic crossing,
and thus produce three images, of which two are very highly magnified
and detectable and the third is not.  Therefore, if the initial GW
detection corresponds to the first of the two highly magnified images,
then the second of the two highly magnified images is generally
detectable.  If the initial GW detection corresponds to the second of
two highly magnified images, then the next appearance corresponds to
the third much less strongly magnified and thus undetectable image.

\begin{figure}[t]
\begin{center}
 \includegraphics[width=120mm]{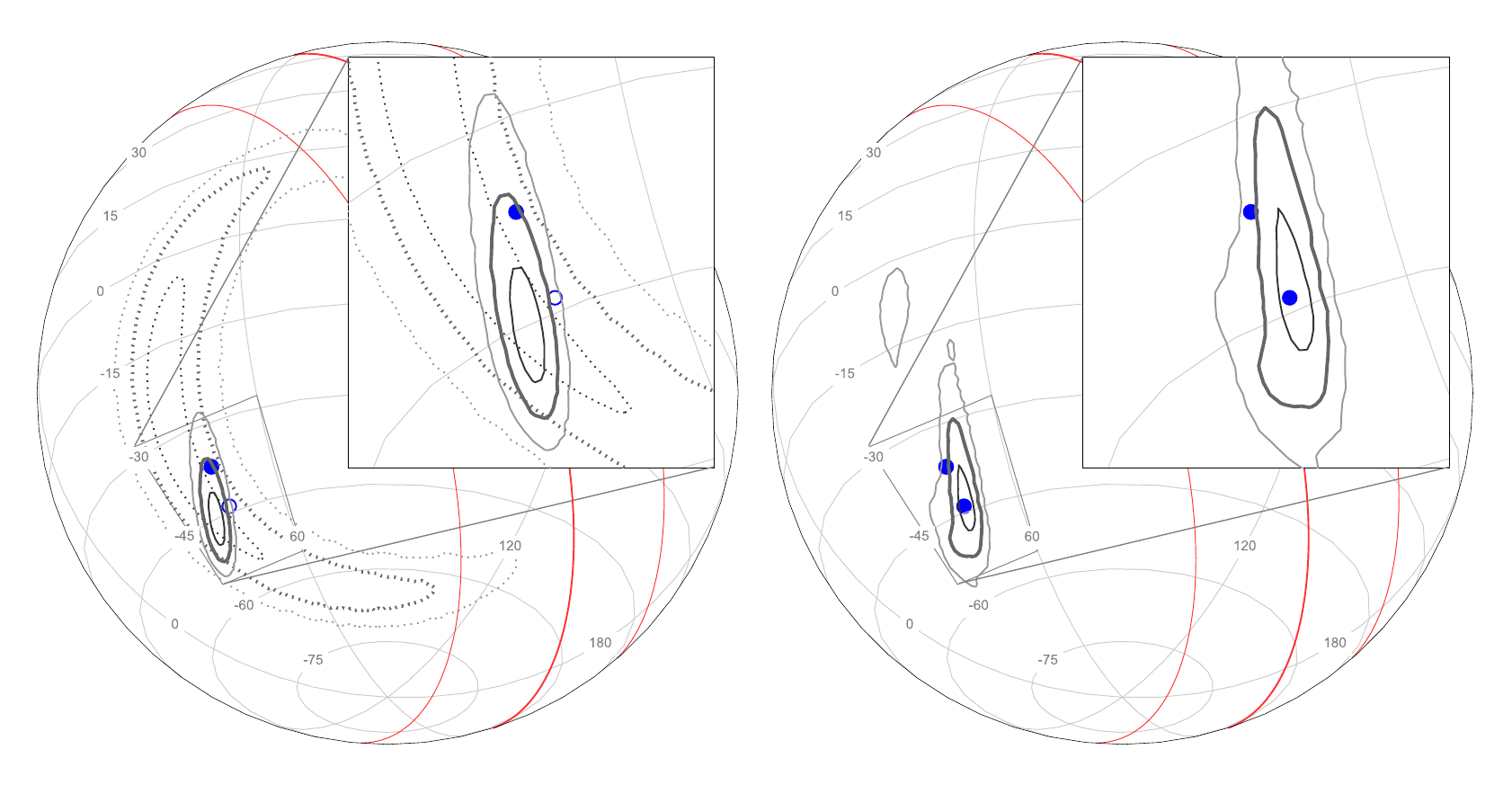} 
 \caption{{\sc Left} -- Initial BAYESTAR skymap for GW170814 from
   GCN21474, showing the LIGO only contours (dotted), joint LIGO-Virgo
   contours (solid), Abell\,3084 (filled blue), and
   SMACS\,J0304.3$-$4401 (open blue).  {\sc Right} -- Initial
   LALInference skymap from GCN21493, showing updated sky localisation
   based on LIGO-Virgo data (solid) and the locations of both
   clusters.  SMACS\,J0304.3$-$4401 is closer to the peak of the sky
   localisation than Abell\,3084 in the revised map.  In both panels
   the 90\% credible region is shown as the thick solid contour, and
   galactic latitudes of $\pm20^\circ$ are shown in red.}
   \label{fig:iair}
\end{center}
\end{figure}

\section{Gravitationally Lensed Gravitational Wave Hunters and O2}

The Gravitationally-lensed Gravitational Wave Hunters collaboration
began observing known strong-lensing clusters located within the 90\%
credible sky localisations of new GW detections in August 2017.  The
aim of these observations was to detect an EM counterpart to a GW
based on the speculation that it had been strongly-lensed by a
cluster.  We used the GMOS instrument on Gemini-South and the MUSE
instrument on VLT to examine the strong-lensing regions of clusters
down to depths of ${\rm AB}\simeq25$ per visit.  For a nominal
gravitational magnification of $|\mu|=100$, we therefore in principle
reach a sensitivity of ${\rm AB}\simeq30$ in the search for an EM
counterpart to GWs.  This is the most sensitive search conducted to
date.  The first visit with each telescope took place as soon as
possible after the LIGO-Virgo alert via a rapid target of opportunity
observation, and subsequent visits took place up to $\sim1$ week
later.

We observed two strong-lensing clusters associated with the sky
localisation of GW170814 (Abbott et al. 2017b), namely Abell\,3084 and
SMACS\,J0304.3$-$4401 (Smith et al. 2017).  Figure~\ref{fig:iair}
shows the location of the clusters and the LIGO-Virgo credible regions
on the celestial sphere.  The small change in the sky localisation
between the initial BAYESTAR and preliniary LALInference skymaps, and
the relationship between these and the cluster coordinates explains
why we initially observed Abell\,3084 and subsequently switched to
SMACS\,J0304.3$-$4401.

\section{Future prospects}

We consider the timescale on which the rate of detection of
strongly-lensed GWs will approach $R_{\rm detect}\simeq1$ per detector
year.  Two factors contribute to this timescale: (1) gains in
sensitivity of the LIGO-Virgo detectors, and (2) expansion of the
sample of known strong-lensing clusters.  LIGO is scheduled to reach
design sensitivity in 2020, by which time it will probe a
$\sim10\times$ larger volume than was the case in O1 and O2 (Abbott et
al., 2017c).  This improved sensitivity will lead to less extreme
magnifications being required to reinterpret GW signals as coming from
redshifts higher than the cluster lenses.  We estimate that the
greater sensitivity and consequent reduced magnification requirement
will together increase the rate to $R_{\rm detect}\simeq10^{-2}$ per
detector year, based on the same population of 130 known strong
lensing clusters discussed above.

Clearly the incidence of strong-lensing of GWs is independent of the
completeness of the available sample of strong-lensing clusters.
However, ability to recognise GWs as being strongly-lensed does depend
on knowledge of the lens population.  Therefore $R_{\rm detect}$ will
grow as more strong-lensing clusters are discovered.  Forecasts for
upcoming large-scale optical/near-infrared surveys indicate that
\emph{Euclid} and LSST will find of order $1$ strong-lensing cluster
per square degree in the redshift range $z\sim0.2-0.5$ considered
above (Boldrin et al., 2016).  It should therefore be possible to
achieve a further two orders of magnitude gain to give $R_{\rm
  detect}\simeq1$ per detector year during the 2020s.  Precisely when
this will be achieved will depend on the observing strategy of each
survey, and the efficiency of strong-lensing cluster detection in the
survey data.  

Additional GW detectors, KAGRA and LIGO-India, are planned to begin
operation in the period 2023-2025 (Abbott et al., 2017c), i.e.\ in
parallel with \emph{Euclid}/LSST discovery of strong-lensing clusters.
These detectors will increase the volume within which GWs can be
detected by a further factor of $3$ beyond the LIGO design
sensitivity.  This will help to increase the rate of strongly-lensed
GW detections in the mid-2020s.

\section*{Acknowledgments}

GPS thanks Iair Arcavi for assistance with producing
Figure~\ref{fig:iair}.  We are grateful for awards of observing time
by the Directors of the European Southern and Gemini Observatories.
We acknowledge financial support from the Science and Technology
Facilities Council and the European Research Council.

\end{document}